\documentclass{article}
\usepackage[T1]{fontenc}
\usepackage[utf8]{inputenc}
\usepackage{ismir} % Remove the "submission" option for camera-ready version
\usepackage{amsmath,cite,url}
\usepackage{color}
\usepackage{siunitx}
\usepackage{tabularray}
\usepackage{booktabs}
\usepackage{graphicx}
\usepackage{multirow}
\usepackage{listings}
\usepackage[table,xcdraw]{xcolor}
\usepackage{algorithm}
\usepackage{algorithmic}
\usepackage{makecell}
\usepackage[bookmarks=false,unicode,hidelinks]{hyperref}

\definecolor{codegreen}{rgb}{0,0.6,0}
\definecolor{codepurple}{rgb}{0.58,0,0.82}
\title{Matchmaker: An Open-Source Library for Real-Time Piano Score Following and Systematic Evaluation}

\multauthor
{Jiyun Park$^{1\ast}$ \hspace{1em} Carlos Cancino-Chac\'on$^{2\ast}$}
{{\bf Suhit Chiruthapudi$^2$ \hspace{1em} Juhan Nam$^1$}\\
$^1$ Graduate School of Culture Technology, KAIST, South Korea\\
$^2$ Institute of Computational Perception, Johannes Kepler University Linz, Austria\\
\tt\small \{june,juhan.nam\}@kaist.ac.kr, \{carlos.cancino\_chacon,suhit.chiruthapudi\}@jku.at}

\def\authorname{J. Park, C. Cancino-Chacón, S. Chiruthapudi and J. Nam}

\begin{document}

\maketitle
\begingroup
\renewcommand\thefootnote{\fnsymbol{footnote}}
\footnotetext[1]{Equal contribution.}
\endgroup
\begin{abstract} 
Real-time music alignment, also known as \emph{score following}, is a fundamental MIR task with a long history and is essential for many interactive applications. Despite its importance, there has not been a unified open framework for comparing models, largely due to the inherent complexity of real-time processing and the language- or system-dependent implementations. In addition, low compatibility with the existing MIR environment has made it difficult to develop benchmarks using large datasets available in recent years. While new studies based on established methods (e.g., dynamic programming, probabilistic models) have emerged, most evaluations compare models only within the same family or on small sets of test data. This paper introduces \emph{Matchmaker}, an open-source Python library for real-time music alignment that is easy to use and compatible with modern MIR libraries. Using this, we systematically compare methods along two dimensions: music representations and alignment methods. We evaluated our approach on a large test set of solo piano music from the (n)ASAP, Batik, and Vienna4x22 datasets with a comprehensive set of metrics to ensure robust assessment. Our work aims to establish a benchmark framework for score-following research while providing a practical tool that developers can easily integrate into their applications.
\end{abstract}

\section{Introduction}\label{sec:introduction}

% \begin{figure}[t]
%   \centering
%   \centerline{\includegraphics[width=\linewidth]{figs/score-following.png}}  
% \caption{An illustrative example of score following}
% %\caption{A task definition of score following}
% \label{fig:score-following}
% \end{figure}

% \begin{figure*}[t]
%   \centering
%   \centerline{\includegraphics[width=\linewidth]{figs/framework.png}}  
% \caption{A conceptual framework of score following}
% \label{fig:framework}
% \end{figure*}

Real-time music alignment, also known as \emph{score following}, is the task of aligning performance data to the corresponding position in the musical score in real-time.
Ever since it was first introduced independently by Roger Dannenberg~\cite{dannenberg_-line_1984} and Barry Vercoe~\cite{vercoe_synthetic_1984} over 40 years ago, music alignment has become one of the fundamental MIR tasks.
Score following is a necessary component of many interactive applications (e.g., automatic accompaniment systems~\cite{cancino-chacon_accompanion_2023,armstrong_real-time_2024,raphael_music_2010,maezawa_i_2024}, automatic page turning\cite{arzt_automatic_2008,henkel_fully_2021}, lyrics alignment or tracking singing voice\cite{brazier_towards_2020,park_real-time_2024,gong_real-time_2015}, audiovisual/multimodal \cite{otsuka_real-time_2011,maezawa_i_2024} and visualizations \cite{lartillot_real-time_2020}.
Music alignment began as real-time score following~\cite{dannenberg_-line_1984,dannenberg_following_1987,dannenberg_new_1988,vercoe_synthetic_1984,vercoe_synthetic_1985,puckette_score_1995} but, by the mid-90s, had diverged into online and offline methods (see, e.g., early offline work by Desain et al. \cite{desain_robust_1997}).

% \cite{cancino-chacon_accompanion_2023,armstrong_real-time_2023,armstrong_real-time_2024,raphael_demonstration_2004,raphael_demonstration_2006,raphael_music_2010,raphael_orchestral_2009,raphael_musical_2004,maezawa_i_2024}, automatic page turning\cite{arzt_automatic_2008,henkel_fully_2021,arzt_score_2008}, lyrics alignment or tracking singing voice\cite{brazier_towards_2020,park_real-time_2024,gong_real-time_2015,brazier_-line_2021,brazier_autonomous_2023,brazier_handling_2021}, audiovisual/multimodal \cite{otsuka_real-time_2011,maezawa_i_2024} and visualizations \cite{lartillot_real-time_2020}.
% Music alignment began as real-time score following~\cite{dannenberg_-line_1984,dannenberg_following_1987,dannenberg_new_1988,vercoe_synthetic_1984,dannenberg_real-time_1989,vercoe_synthetic_1985,puckette_score_1992,puckette_explode_1990,puckette_score_1995} but, by the mid-90s, had diverged into online and offline methods (see, e.g., early offline work by Desain et al. \cite{desain_robust_1997}).

From its early use on monophonic sources like voice\cite{puckette_score_1995} and wind instruments, score following has grown to support polyphonic instruments such as piano, ensemble, and even full orchestral performances~\cite{puckette_score_1995,raphael_orchestral_2009,prockup_orchestral_2013,arzt_real-time_2015}. Research has also expanded across input modalities of the performance, with systems operating on audio or MIDI, and score representations including string format, symbolic score, and sheet image\cite{henkel_score_2019}.
% Already in the early 90s, score following was applied in music production and sequencing software \cite{puckette_explode_1990,puckette_score_1992}.
% At IRCAM, Puckette and colleagues used score following in concerts to synchronize live electronics with instruments \cite{puckette_score_1992}, though, as noted by Puckette and Lippe, \emph{“algorithms that work well in theory, or even in the laboratory, often cease working when confronted with real examples of instrument writing”} \cite{puckette_score_1992}.

% Real-time music alignment, commonly referred to as \emph{score following}, is a foundational task in music information retrieval (MIR) with broad use-cases in many interactive real-world applications such as page turner, live accompaniment system. 
% The problem was first introduced in its online version in 1984 with the seminal work by Roger Dannenberg \cite{dannenberg_-line_1984} and Barry Vercoe \cite{vercoe_synthetic_1984}. 
% Ever since, score following has evolved into a key technology bridging symbolic music representations with real-time acoustic performance.

The score following challenge~\cite{cont_evaluation_2007} in MIREX laid the foundation to formalize the evaluation framework, introducing important metrics that include considerations in real-time. 
However, many subsequent studies have been developed in different environments—ranging from system-dependent~\cite{cont_antescofo_2008,echeveste_improved_2023} to language-dependent~\cite{dixon_match_2005, joder_learning_2013} implementations—often tailored to specific use cases and without publicly shared source code.
As a result, implementations became fragmented across platforms, making it difficult to extend, reproduce, or compare methods in a unified setting.
This has hindered the development of a unified evaluation framework and comparison over methods or features on shared datasets remain rare, limiting the generalizability and reproducibility.

In this paper, we address these challenges by proposing 
a unified, open framework for the evaluation and benchmarking of real-time audio-based score following.
Considering public datasets that offer a range of difficulty levels, multiple renditions, and precise beat-level annotations, we base our evaluation on three representative piano performance datasets.
We implement this framework as an open-source Python package called \emph{Matchmaker},\footnote{\url{https://github.com/pymatchmaker/matchmaker}} that allows real-time execution of representative baselines of score following algorithms. %through a user-friendly interface.
In addition to benchmarking, it supports audio device input and has been validated in application contexts through a standalone demo system.

\section{A Conceptual Framework for Score Following}\label{sec:framework}

As a way to organize and compare the components of systems for score following, we follow the structure proposed by Müller \cite{muller_towards_2004}.
This framework consists of three core components: (1) input music representations, (2) features, and (3) online alignment algorithms.

\subsection{Music Representation}

% Real-time score following requires a pair of temporally aligned data streams: a fixed reference derived from the score, and a time-evolving input derived from the live performance. 
% The score can take various symbolic formats (e.g., MIDI, MusicXML) or sheet images, and is typically converted into an intermediate representation such as synthesized audio or event sequences for comparison.

% The performance input may be given as either audio or MIDI, each with distinct representational and computational characteristics. 
% Audio input is continuous and high-dimensional, requiring strict latency constraints. 
% In contrast, MIDI input is discrete and event-based, making it easier to interpret in terms of note onsets and pitches, though less representative of expressive timing and dynamics in acoustic performance.

% Performance signals also vary in musical and acoustic complexity. 
% Some instruments, such as piano or guitar, produce discrete pitches and are often polyphonic, while others like violin, voice, or flute exhibit continuous pitch variation and require different handling in alignment. 
% Furthermore, multi-instrument recordings introduce timbral overlap and source separation challenges, especially in audio-based systems. 
% These representational differences must be accounted for when designing alignment algorithms and evaluation protocols.

Score following aligns a fixed reference derived from musical scores with a time-evolving input from a performance. 
The score can take various symbolic formats (e.g., MIDI, MusicXML) or sheet images, and is typically converted into an intermediate representation such as synthesized audio or event sequences.
The performance input may be given as either audio or MIDI, each with distinct representational and computational characteristics. 
Audio input is continuous and latency-sensitive, while MIDI is discrete and event-based. 
Instrumental factors also affect alignment design: polyphonic or discrete-pitch instruments (e.g., piano) differ from continuous-pitch sources (e.g., violin, voice). Multi-instrument recordings pose further challenges due to timbral overlap and source ambiguity.

\subsection{Features}

% The performance and the reference score are typically transformed into a shared feature space that captures pitch-related information while remaining robust to expressive variation and tempo fluctuation.
% The most common feature used in music synchronization is chroma features~\cite{duan_state_2011, chou_simple_2018,muller_music_2021} with lots of variants\cite{perez2022comparison} for computing them.
Chroma features are the most commonly used in music synchronization, with many variants for their computation~\cite{duan_state_2011, chou_simple_2018, muller_music_2021, perez2022comparison}.
Other works also use various spectral features such as constant-Q transforms (CQT)~\cite{joder_learning_2013, chen_efficient_2016}, non-negative matrix factorization(NMF)-based~\cite{carabias-orti_audio_2015} or spectral template~\cite{korzeniowski_refined_2013} for improved polyphonic alignment.
Beyond spectral representations, context-aware features such as onset-based feature \cite{ewert2009high} or beat-synchronous frames have been introduced to capture temporally salient events useful for alignment.
Later work explored learned features, including feedforward mappings~\cite{joder_learning_2013}, semi-supervised decompositions like NMF, and more recent neural approaches~\cite{pillay_neural_2024}. 
% While these offer richer contextual information, they often rely on fixed-length inputs and introduce latency, making real-time use more challenging.
% Many works also explored learned feature representations, aiming to extract alignment-relevant representations directly from data. Early studies focused on feedforward mappings to score-aligned templates~\cite{joder_learning_2013}, while others employed semi-supervised decomposition such as NMF. 
% More recent studies introduce ~\cite{pillay_neural_2024}
% % while later systems leveraged convolutional attention~\cite{agrawal_convolutional-attentional_2022} and structure-aware modeling~\cite{agrawal_structure-aware_2021}. 
% These features offer richer contextual and timbral sensitivity compared to off-the-shelf or handcrafted ones, but often require input windows constraints and substantial training data—making real-time application challenging due to latency and generalization concerns.
While these offer richer contextual information, they often rely on fixed-length inputs and introduce latency, making real-time usage more challenging.

% Spectral features:
% \begin{itemize}
% \item korzeniowski \cite{korzeniowski_refined_2013}
% \item Chromagram \cite{duan_state_2011,chou_simple_2018}
% \item Constant-Q Transform \cite{joder_learning_2013,chen_efficient_2016}
% \end{itemize}
%  Learned features:
%  \begin{itemize}
%  \item Joder: \cite{joder_learning_2013,joder_improved_2010} 
%  \item Agrawal \cite{agrawal_convolutional-attentional_2022,agrawal_learning_2020,agrawal_towards_2022,agrawal_hybrid_2019,agrawal_structure-aware_2021}
% \end{itemize}

\begin{figure*}[t]
  \centering
  \centerline{\includegraphics[width=\linewidth]{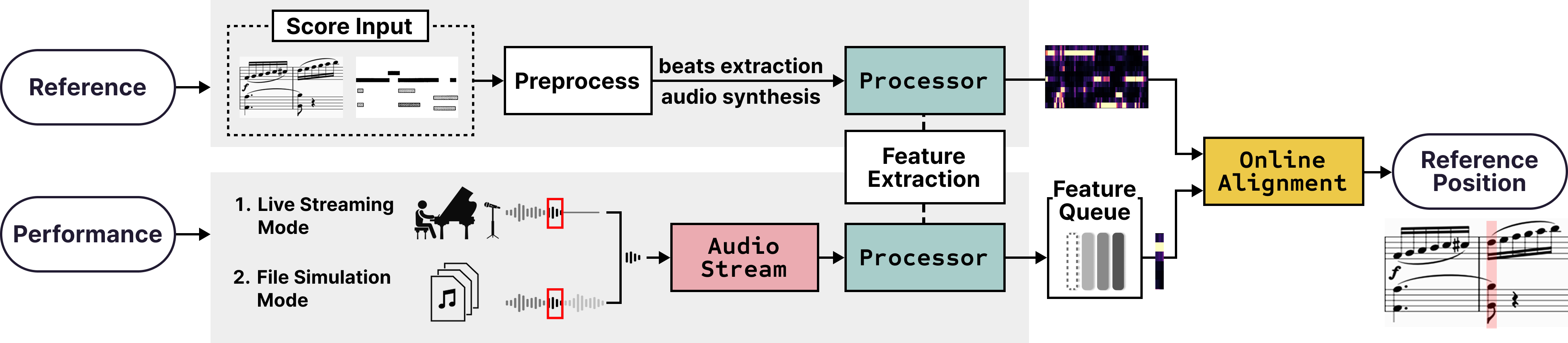}}
\caption{Overview of the score following package}
\label{fig:overview}
\end{figure*}

\subsection{Alignment Algorithms}

Two major families of alignment algorithms have been used in score following: dynamic programming and probabilistic models.

The dynamic programming approach, especially dynamic time warping (DTW), aligns two sequences by minimizing cumulative cost. Its online variant, On-Line Time Warping (OLTW)~\cite{dixon_-line_2005}, enables causal alignment within a fixed-size of window. Variants include windowed~\cite{macrae_polyphonic_2008}, parallel~\cite{rodriguez-serrano_real-time_2015}, and constrained DTW~\cite{carabias_real-time_2012, rodriguez-serrano_real-time_2015}, as well as tempo-aware extensions~\cite{arzt_real-time_2010,arzt_real-time_2015}.

Probabilistic state-space models offer an alternative by treating alignment as latent state inference under uncertainty~\cite{cano_score-performance_1999, cont_antescofo_2008, duan_state_2011}. HMM-based systems model each note as a sequence of states (e.g., attack–steady–release), with extensions including semi-Markov~\cite{nakamura_autoregressive_2015}, hybrid~\cite{raphael_orchestral_2009}, and Bayesian variants~\cite{raphael_bayesian_2001}. Kalman filter models and switching state-space systems~\cite{yamamoto_real-time_2012, jiang_score_2020} further incorporate tempo dynamics, while particle filters~\cite{duan_state_2011, otsuka_real-time_2011} handle multimodal uncertainty in real time.

Other paradigms include early string-matching algorithms~\cite{dannenberg_-line_1984} and reinforcement learning-based approaches for multimodal or visual score alignment~\cite{dorfer_learning_2018}.

\section{Implementation}\label{sec:matchmaker}

\subsection{Python Package Structure}

\emph{Matchmaker} is an open source Python package that implements representative real-time music alignment algorithms within a modular, extensible framework.
Figure \ref{fig:overview} illustrates the overview of the package and the whole pipeline. 
The current version of \emph{Matchmaker} provides two types of algorithms: 1) online time warping, with two variants: \texttt{OLTWDixon}, based on the methods proposed in \cite{dixon_-line_2005,dixon_live_2005}, and \texttt{OLTWArzt}, based on \cite{arzt_towards_2010, arzt_real-time_2015}; and 2) an HMM-based algorithm, similar to the one used in \cite{cancino-chacon_accompanion_2023,jiang_score_2020}. 
 A full description of the algorithms and their parameters can be found in the supplementary Appendix.\footnote{\url{https://pymatchmaker.github.io/ismir2025_supplementary_materials/}}

\emph{Matchmaker} supports two main usage scenarios: (1) live streaming mode using the audio device and (2) simulation mode, which processes a performance file as input. Figure \ref{fig:matchmaker_example} shows an example of running live streaming mode with the default setting. 
The \texttt{AudioStream} object handles the input stream by chunking the audio with overlapping windows to avoid padding artifacts. 
% Real-time input is managed using PyAudio\footnote{\url{https://people.csail.mit.edu/hubert/pyaudio/}} as the backend.
Both the synthesized score audio and the performance audio are passed to a \texttt{Processor} object that performs feature extraction. The extracted features are pushed into a queue and consumed by the \texttt{OnlineAlignment} object, which runs the alignment methods in real time.
\emph{Matchmaker} takes a musical score with all symbolic music formats (MusicXML, MIDI, MEI, etc.) available by \textit{partitura}.\footnote{\url{https://github.com/CPJKU/partitura}}
The returned output is the current position in the score, represented in beats as a musical unit according to the time signature in the piece. 
More detailed description and API documentation of the package are available here.\footnote{\url{https://pymatchmaker.readthedocs.io/}}

\subsection{Design and Implementation Details}

\begin{figure}[t]
\begin{lstlisting}[language=Python]
from matchmaker import Matchmaker

mm = Matchmaker(
    score_file="path/to/score.musicxml",
    input_type="audio",
)
for current_position in mm.run():
    print(current_position)
\end{lstlisting}
\caption{A code example for running the \emph{Matchmaker} in a live streaming mode.}\label{fig:matchmaker_example}
\end{figure}

We provide a simple and user-friendly interface to run the score following with minimal setup. 
As shown in Figure \ref{fig:matchmaker_example}, users can instantiate a \texttt{Matchmaker} object with a score file and execute a run that iterates over the estimated score position for each step.
To streamline real-time processing, the \texttt{AudioStream} class is implemented as a context manager that automatically handles stream initialization and teardown. 
Furthermore, the alignment process is designed as a generator, enabling users to receive score positions concurrently while the alignment is in progress. 
This design allows for efficient real-time integration without requiring users to manage multiple threads, buffers, or callbacks explicitly.

While the online mode uses a multi-threaded queue for asynchronous audio buffering, the simulation mode processes audio chunks in advance within a single-threaded setup. By decoupling real-time I/O concerns from core alignment evaluation, it is intended to avoid variability from Python version, OS-level threading, or queuing delays, ensuring a consistent and reproducible benchmarking environment.
In addition, OLTWArzt is implemented in Cython \cite{behnel2011cython} for efficiency, a superset of Python designed for C-like performance by incorporating C data types and optimizing the execution of Python code.

\section{Experiments}\label{sec:experiments}

\subsection{Datasets}\label{sec:datasets}

% Statistics - # Composer, # Performances, difficulty level, durations, total beat events, etc.
We use three public piano performance datasets: (n)ASAP \cite{peter_automatic_2023}, Batik \cite{hu2023batik} and Vienna 4x22 \cite{vienna4x22}, each of them offering complementary characteristics for benchmarking score following.
% These datasets were chosen for their varying levels of performance expressivity, stylistic diversity, and quality of annotation. 
(n)ASAP, a subset of the MAESTRO dataset including note-level score alignments, includes expressive performances of technically demanding solo piano pieces, offering high difficulty and stylistic diversity. We use only the pieces in the MAESTRO v2 test split.
Vienna4x22 provides 22 distinct renditions for each of four relatively easy pieces, which is suitable to test robustness to interpretive variation.
Batik dataset contains recordings of 12 Mozart sonatas by a single pianist with  the longest average piece duration among the three datasets, enabling evaluation across long-form classical repertoire.

We use ground-truth beat-level annotations provided with the (n)ASAP dataset, and extract equivalent annotations for Batik and Vienna4x22 from the \textit{.match} files~\cite{cancino2022match}, which contain note-wise score–performance alignments. 
%For generating \textit{.match} files, we use the \texttt{parangonar} tool~\cite{peter-offline2023} to perform MIDI-to-score alignment based on MusicXML scores.
% Although all the datasets provides precise level of alignment, it is not directly applicable to extract consistent beat annotations that matches 
In addition, we incorporate the difficulty levels of each piece based on G. Henle Publishers,\footnote{\url{https://www.henle.de/Levels-of-Difficulty/}} which provides a 1-to-9 grading scale. The pieces used in our experiments span levels 4 through 9, representing a diverse set of works above intermediate level.
Table~\ref{tab:dataset-overview} provides the detailed statistics of the datasets.
% pianolibrary.org.\footnote{https://www.pianolibrary.org/about/difficulty-scale/}

We only included performances in the experiment that recorded an MAE of less than 100 ms in the offline test, using the \textit{synctoolbox}\footnote{\url{https://github.com/meinardmueller/synctoolbox}} with Chroma \& DLNCO features.
The evaluation was conducted on 184 performances across 93 pieces, totaling over 58,000 beats and 247,000 notes, with an overall duration of 7.74 hours of performances and a piece-wise average difficulty of 6.11.

\begin{table}[t]
\centering
\footnotesize
\renewcommand{\arraystretch}{1.2}
\setlength{\tabcolsep}{4.5pt}
\begin{tabular}{lcccccc}
\toprule
\textbf{Dataset} & \textbf{\#Pieces} & \textbf{\#Perf} & \textbf{\#Beats} & \textbf{\#Notes} & \textbf{Dur (h)} & \textbf{Difficulty} \\
\midrule
(n)ASAP     & 43 & 59  & 26,329  & 100,958  & 2.65 & 6.53 \\
Batik    & 30 & 30  & 18,789  & 102,421  & 2.85 & 5.67 \\
Vienna   &  4 & 88  & 13,728  &  43,656  & 2.24 & 4.88 \\
\midrule
\textbf{Total} & \textbf{77} & \textbf{177} & \textbf{58,846} & \textbf{247,035} & \textbf{7.74} & \textbf{6.11} \\
\bottomrule
\end{tabular}
\caption{Datasets used in the evaluation.}
\label{tab:dataset-overview}
\end{table}

\subsection{Experiment Settings}

We conducted all evaluations under simulation-based conditions to ensure reproducibility. 
Live testing was avoided due to variability introduced by room acoustics and hardware setup, which complicates fair comparison across systems. 
The accuracy tests were carried out on an Intel i9-9900K CPU (16 cores @ \SI{3.6}{\giga\hertz}), Python 3.9, with a sample rate of 44.1\,kHz and a frame rate of 30, chosen to balance latency and alignment accuracy.
We tested chromagram, mel-spectrogram, constant-Q transform (CQT), mel-frequency cepstral coefficients (MFCCs)~\cite{brazier_addressing_2020} and a simple STFT-based onset-sensitive representation similar to the one used in Dixon~\cite{dixon_-line_2005}, which we name log-spectral energy (LSE). 
While results for all features were evaluated, we report detailed latency and accuracy metrics for the best-performing configuration of each model.
To account for hardware variability, latency was measured in multiple setups: an Intel i9-9900K, an Apple M4 MacMini, and an Apple M2 Pro MacBook, with the reported latency values averaged across these devices.

\subsection{Preprocessing}

In the preprocessing step (see Fig. \ref{fig:overview}), the symbolic scores are synthesized to audio using FluidSynth, provided by \textit{partitura}. 
Since MusicXML often lacks tempo markings, we set the synthesis tempo to each performance’s average—rounded to the nearest 20 BPM—assuming performers follow approximate tempo indications.
% Since MusicXML often lacks tempo markings, we loosely adjusted the tempo to match the performance average, rounding to the nearest 20 BPM under the assumption that performers follow approximate tempo indications.

To generate beat annotations for the synthesized score audio, we computed beat positions using the synthesis tempo and the score’s time signature. 
For compound meters (e.g., 6/8, 9/8, and 12/8), we adopted (n)ASAP’s beat annotation rules—counting them as two, three, and four beats per measure, respectively—across all datasets to align score-side annotations with performance annotations.
Based on the synthesized audio, we then extract the feature using the same \texttt{Processor} used in the online phase, but precompute them offline for the entire score sequence.
% For compound meters which typically , we refer to (n)ASAP's beat annotation rules (e.g., counting \setmeter{6}{8} meters as two beats per measure) across all datasets to align score‐side annotations with their performance annotations.

\section{Evaluation}\label{sec:evaluation}

\begin{figure}[t]
  \centering
  \centerline{\includegraphics[width=\linewidth]{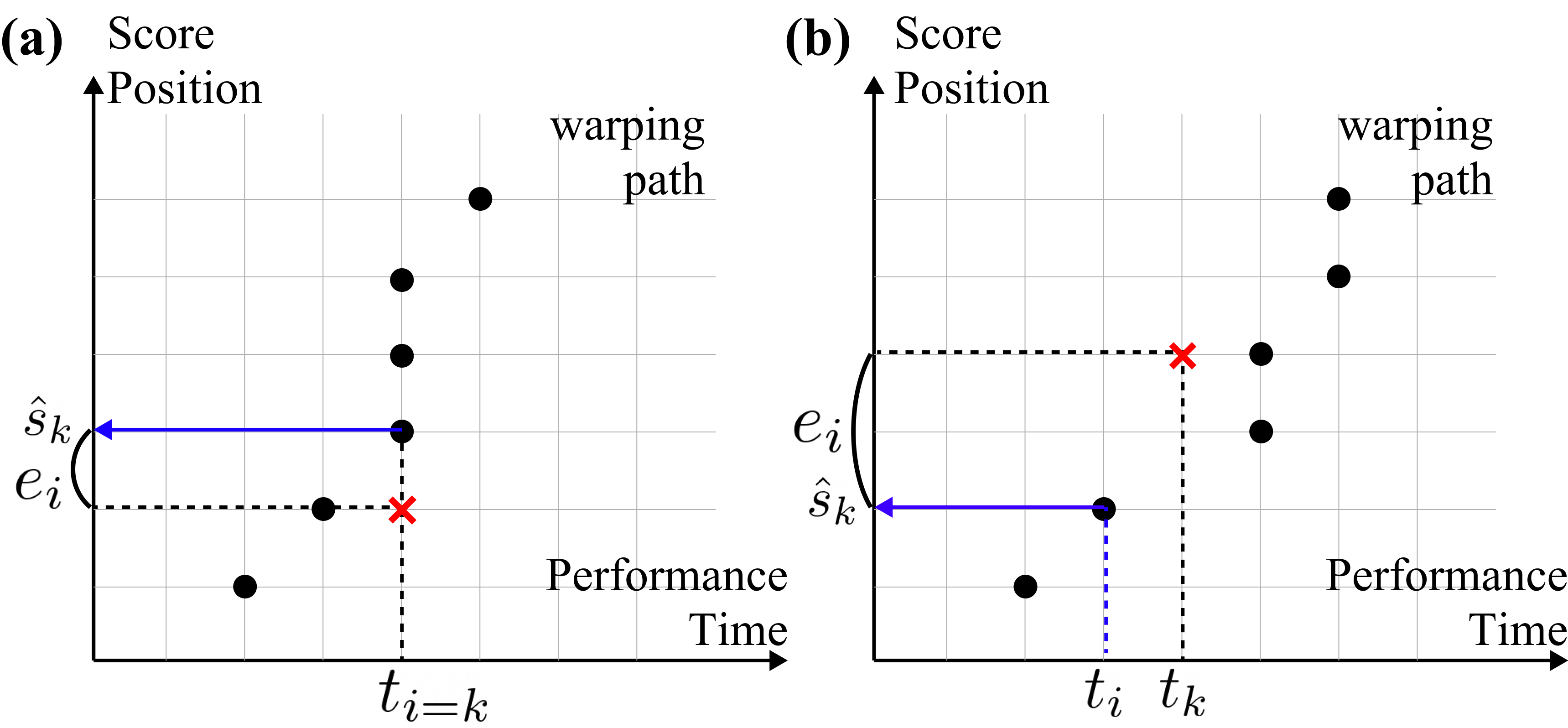}}  
\caption{Two examples of error calculation using the mapping function. (a) shows a one-to-many alignment at the evaluation point, while (b) illustrates a skipped alignment.}
\label{fig:mapping_ex}
\end{figure}

Evaluating score following is challenging due to causality, timing precision, and output latency. Since the MIREX challenge~\cite{cont_evaluation_2007} provided foundational metrics, later studies introduced alternative evaluation strategies including beat-level evaluations or asynchrony~\cite{cancino-chacon_accompanion_2023}, reflecting the task's frequent integration with automatic accompaniment systems. 

In this work, we adopt two complementary evaluation perspectives. First, we evaluate in the performance domain, where errors are measured in milliseconds based on ground-truth annotations aligned to the audio. This approach is commonly used in audio-to-score alignment research and enables precise, frame-level evaluation, since the annotations directly reflect the actual timing of the performance.
Second, we also evaluate in the score domain measured in beat units as suggested in \cite{duan_state_2011,morsi2022bottlenecks}, which better reflects the nature of score following as a task of predicting the corresponding score position at each moment of the performance. 
% Since score is represented as symbolic data, the beat positions can be clearly defined.

\subsection{Evaluation Metrics}

\begin{figure}[t]
  \centering
  \centerline{\includegraphics[width=\linewidth]{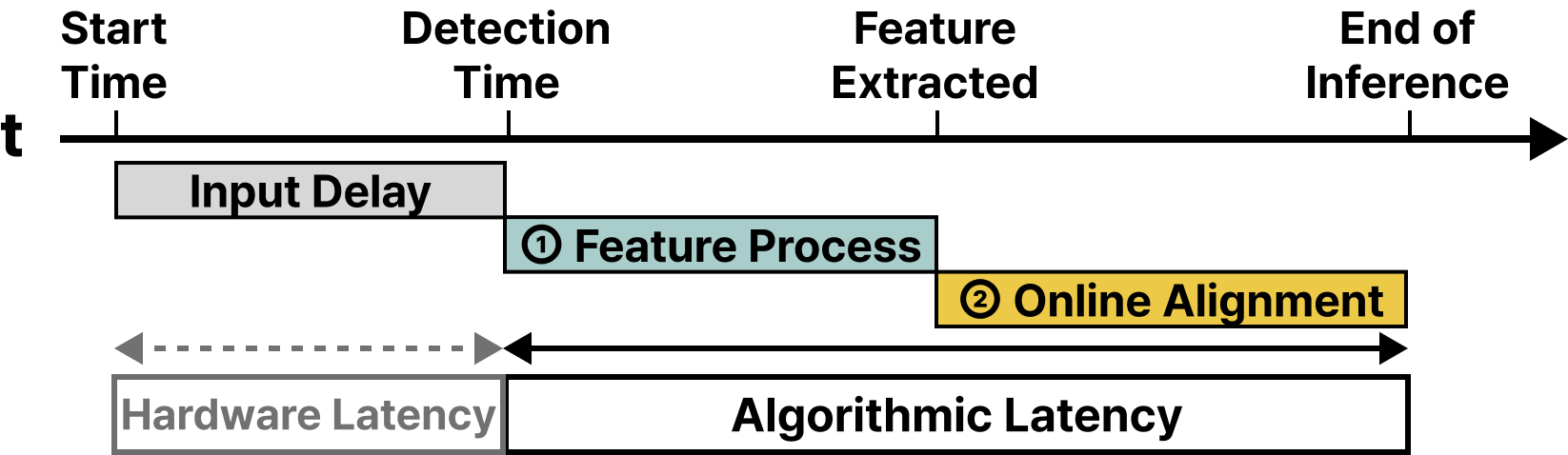}}  
\caption{Defined delay types of the system. Only system delay is considered in the experiment.}
\label{fig:delay}
\end{figure}

We select evaluation metrics mostly adapted from score following MIREX benchmark \cite{cont_evaluation_2007} and audio-to-score alignment (ASA) metrics \cite{morsi2022bottlenecks}.
% We use \textbf{Alignment Rate} (AR) with a tolerance range of \(|\theta_e|\) from 50ms to 2000ms, Absolute Errors (AE) in milliseconds and in beats which derives \textbf{Average Absolute error} (AAE), \textbf{Median Absolute Error} (MAE) with variance \(\sigma_e\), kurtosis and skewness, and \textbf{average latency} \(\mu_l\). 
We use \textbf{Alignment Rate} (AR) within a tolerance range of $|\theta_e|$, varying from 50\,ms to 2000\,ms. We also compute \textbf{Absolute Errors} (AE), both in milliseconds and in beats, from which we derive the \textbf{Average Absolute Error} (AAE) and \textbf{Median Absolute Error} (MAE), along with the standard deviation $\sigma_e$.
To further characterize the distribution of errors, we report \textbf{kurtosis} and \textbf{skewness} which capture the peakedness and asymmetry of the non-absolute error distribution, respectively. 
% In addition, we include the \textbf{average latency} $\mu_{lat}$.
In addition, we report the \textbf{average latency} $\mu_{\mathrm{lat}}$, defined as the \textit{system delay} from the detection time to the end of inference. 
Unlike total latency, this excludes \textit{hardware latency} and is composed of two parts: (i) feature processing and (ii) execution of the online alignment algorithm for each frame step (see Fig.~\ref{fig:delay}).
Errors exceeding 2 seconds (or 2 beats in the score domain) are excluded from AE calculations, including both AAE and MAE, to avoid distortion from unbounded tracking failures.
We report AR in two ways. The averaged piece-wise AR is a common measure, while the total AR reflects the proportion of successfully aligned beat events across the entire dataset. The latter avoids overrepresentation of shorter pieces and provides a more balanced view of overall performance.

% As tracking failures can cause alignment errors to increase drastically, we report MAE alongside AAE as a more reliable measure of alignment accuracy under outlier conditions.
% We designed our framework that enables multiple levels of evaluation (note-, beat-, measure-) depending on the available annotation levels of the potential datasets, here we use beat-level evaluation according to \cite{morsi2022bottlenecks} and additionally report the AE normalized in beats. 

To evaluate runtime latency under simulation, we measure two components: the average duration for extracting features from incoming audio frames, and the time taken by the alignment process to consume features and predict score positions. Specifically, the latency was computed from the moment audio was read to the time the score position was predicted—excluding hardware I/O delays. This two-step measurement allows for standardized latency reporting independent of the hardware setup. 

\begin{table*}[!t]
\centering
\footnotesize
\setlength{\tabcolsep}{4pt}
\renewcommand{\arraystretch}{1.2}
\begin{tabular}{llcccccccccc}
\toprule
\multirow{2}{*}{\textbf{Dataset}} & \multirow{2}{*}{\textbf{Method}} &
\multirow{2}{*}{\textbf{AAE(ms)↓ ± $\sigma$}} & \multirow{2}{*}{\textbf{MAE(ms)↓}} &
\multirow{2}{*}{\textbf{Skew.}} & \multirow{2}{*}{\textbf{Kurt.}} &
\multicolumn{5}{c}{\textbf{Piece-wise AR (\%) ↑}} &
\multirow{2}{*}{\makecell{\textbf{Total AR↑} \\ ($\leq$2000ms, \%)}} \\
\cmidrule(lr){7-11}
& & & & & & $\leq$50ms & $\leq$100ms & $\leq$500ms & $\leq$1000ms & $\leq$2000ms & \\
\midrule
(n)ASAP & OLTWDixon & 189.55 ± 281.55 & 97.09  & 3.20  & 17.97  & 40.3 & 58.5 & 82.5 & 88.3 & 92.0 & 89.4 \\
     & OLTWArzt  & 183.56 ± 263.95 & \textbf{91.18}  & 0.75  & 11.79  & 44.1 & 58.3 & 84.8 & 92.0 & 95.1 & \textbf{92.8} \\
     & HMM      & 487.73 ± 423.27   & 346.01  & 0.18 & 3.33  & 15.6 & 22.2 & 37.5 & 43.8 & 43.8 & 43.8 \\
\midrule
Batik & OLTWDixon & 186.97 ± 262.55 & \underline{104.40} & 3.75  & 24.70  & 28.2 & 51.7 & 82.1 & 85.2 & 87.6 & \underline{89.4} \\
      & OLTWArzt  & 193.36 ± 269.13 & \underline{107.15} & 1.00  & 12.63  & 35.9 & 53.0 & 82.2 & 87.4 & 90.3 & \underline{89.7} \\
      & HMM      & 693.63 ± 376.58 & 641.77 & 0.11  & 0.98   & 4.5  & 10.8 & 34.0 & 46.2 & 64.2 & 61.9 \\
\midrule
Vienna4x22 & OLTWDixon & 285.43 ± 390.82 & \textbf{132.73} & 1.57 & 5.90  & 26.6 & 43.2 & 72.4 & 80.0 & 85.5 & 82.5 \\
           & OLTWArzt  & 300.41 ± 368.70 & 152.51 & 0.50 & 3.93  & 33.2 & 44.5 & 73.3 & 84.3 & 86.7 & \textbf{86.7} \\
           & HMM      & 439.64 ± 427.02   & 319.13  & 0.15 & 3.79 & 23.5 & 33.3 & 51.1 & 57.1 & 63.0 & 75.9 \\
\bottomrule
\end{tabular}
\caption{Evaluation results on three datasets using different score-following methods. The piece-wise alignment rate (AR) is measured as the average over pieces, while the total AR indicates the global proportion of aligned beat events across the entire dataset. All tests were conducted with STFT-based Chroma as features.
}
\label{tab:evaluation-results}
\end{table*}

\subsection{Alignment Mapping Function}\label{sec:mapping-function}

Given the alignment path, the alignment mapping function is applied to transfer the beat positions on one axis (either performance or score) to another axis to compute the alignment error.
Due to the local, stepwise nature of real-time alignment, the resulting path is not necessarily monotonic and may contain multiple correspondents or skipped positions, depending on the implementation and purpose of the methods.
Unlike linear interpolation methods commonly used in offline audio-to-score alignment, which assume continuous mappings, our evaluation relies only on predictions made prior to or at each evaluation time point.
To reflect this, we define the mapping function as follows:
\[
\hat{u}_k = \min\bigl\{\,u_i \mid (u_i, v_i)\in\mathcal{W},\; v_i = \max\{v_j \mid v_j \le k\}\bigr\},
\]
where \(\mathcal{W} = \{(u_i, v_i)\}\) is the warping path expressed in the frame indices: \(u_i\) is the score‐rendered‐audio frame index and \(v_i\) is the performance‐audio frame index.  The inner \(\max\) finds the latest performance frame \(v_i\) not exceeding the current frame \(k\), and the outer \(\min\) selects the smallest score frame \(u_i\) among those alignments.  
This mapping relies solely on past or current frames to maintain causality. It handles skipped or one‐to‐many mappings and avoids any interpolation methods that depend on future frames.

% To reflect this, we define the mapping function as follows:
% \begin{equation*}
% \hat{s}_k = \min \left\{ s_i \mid (s_i, t_i)\in\mathcal{W},\ t_i = \max \{ t_j \mid t_j \leq t_k \} \right\}
% % \hat{s}_k = \min \{s_i \mid (s_i, t_i) \in \mathcal{W},\ t_i = \max\{t_j \mid t_j \leq \tau_k\} \}
% \end{equation*}

% where \( \hat{s}_k \) is the predicted score position for the performance beat \( t_k \), and \( \text{wp} = \{(s_i, t_i)\} \) denotes the warping path consisting of aligned score and performance positions.
% It first identifies the latest performance time \( t_i\) given \( t_k\) from the warping path, then selects the minimum score position \( s_i \) among all alignments at that time.
% This formula accounts for skipped alignments and one-to-many mappings, while ensuring that the evaluation remains causal.

\section{Results}

Table~\ref{tab:evaluation-results} presents a comparison of alignment methods based on performance-domain evaluation, measured in milliseconds.
All methods exhibit positive skewness in error distribution, reflecting the expected lag of the beat estimates in real-time alignment.
The overall results show that the OLTW-based method outperforms the HMM baseline across all datasets in both alignment accuracy and coverage. 
While OLTWDixon and OLTWArzt show comparable MAE depending on the dataset, OLTWArzt consistently achieves higher coverage (\textit{Total AR}), suggesting that it is more robust against overall failures.
The difference likely stems from OLTWDixon skipping uncertain regions, while OLTWArzt’s ``backward-forward'' strategy corrects early misalignments and enhances coverage.
Despite having the lowest AR, the HMM shows the lowest skewness and kurtosis primarily because significant errors (>2 s) are excluded from the summary statistics and its ``sticky” behavior to linger in the same state in local regions tends to narrow the error distribution.

% This is further supported by the skewness values, which are notably lower for OLTWArzt, implying that its error distribution is more symmetric and less prone to large outliers. 
% Overall, the results suggest that OLTWDixon may favor precision in correctly tracked events, whereas OLTWArzt offers better resilience against alignment failures.

% \subsection{Score-Domain Evaluation}

\begin{table}[t]
\centering
\footnotesize
\renewcommand{\arraystretch}{1.1}
\setlength{\tabcolsep}{3pt}
\begin{tabular}{llccc}
\toprule
\textbf{Dataset} & \textbf{Method} & \textbf{AAE↓(beats) ± $\sigma$} & \textbf{MAE↓(beats)} & \textbf{AR↑ (\%)} \\
\midrule
(n)ASAP    & OLTWDixon  & 0.22 ± 0.27 & 0.13 & 83.4 \\
           & OLTWArzt   & 0.27 ± 0.30 & 0.16 & 85.2 \\
           & HMM        & 0.80 ± 0.54 & 0.66 & 76.9  \\
\midrule
Batik      & OLTWDixon  & 0.20 ± 0.27 & 0.11 & 88.9 \\
           & OLTWArzt   & 0.29 ± 0.34 & 0.18 & 88.8 \\
           & HMM        & 0.80 ± 0.38 & 0.67 & 59.3  \\
\midrule
Vienna4x22 & OLTWDixon  & 0.31 ± 0.33 & 0.19 & 78.3 \\
           & OLTWArzt   & 0.37 ± 0.38 & 0.24 & 84.0 \\
           & HMM        & 0.76 ± 0.78 & 0.51 & 70.3  \\
\bottomrule
\end{tabular}
\caption{Beat-level evaluation results including total alignment rate (AR) (\%). }
\label{tab:beat-eval}
\end{table}

\begin{table}[t]
\centering
\footnotesize
\renewcommand{\arraystretch}{1.1}
\setlength{\tabcolsep}{4pt}
\begin{tabular}{lcc|lc}
\toprule
\multicolumn{3}{c|}{\textbf{Feature Process}} & \multicolumn{2}{c}{\textbf{Online Alignment}} \\
\cmidrule(r){1-3} \cmidrule(l){4-5}
\textbf{Type} & \textbf{MAE (ms)} & \textbf{Latency (ms)} & \textbf{Method} & \textbf{Latency (ms)} \\
\midrule
Chroma        & 265.50 & 3.05  & OLTWDixon  & 1.22 \\
mel           & 297.92 & 3.40  & OLTWArzt   & 0.07 \\
CQT           & 341.25 & 42.58 & HMM       & 3.59 \\
LSE           & 241.85 & 0.91  &  \\
MFCC          & 931.81 & 2.58  &  \\   
\bottomrule
\end{tabular}
\caption{Comparison of feature types and alignment methods in terms of alignment error (MAE) and latency. LSE is log-spectral energy feature that was adopted in \cite{dixon_-line_2005}. Latency values are averaged over the hardware setups evaluated in Section \ref{sec:experiments}.}
\label{tab:feature-delay}
\end{table}

Table~\ref{tab:beat-eval} presents an evaluation comparison in beat units, offering a tempo-normalized perspective. The overall trend mirrors the performance-domain results in millisecond, but these results are standardized across tempi. AAE remains around 0.3 beats, with median values typically below 0.2. Total AR is consistently lower than the \SI{2000}{\milli\second}-based metric, reflecting that most pieces have tempi above 60BPM, where two beats span less than two seconds.

In addition, a comparison of various feature types and latencies of the alignment methods are reported in Table~\ref{tab:feature-delay}. Among the features, log-spectral energy (LSE) shows the lowest MAE (\SI{241.85}{\milli\second}) and delay (\SI{0.91}{\milli\second}), indicating strong performance with minimal overhead. In contrast, CQT and MFCC yield higher MAE, with CQT also requiring considerable extraction time (\SI{42.58}{\milli\second}), which limits its real-time suitability. For alignment methods, OLTWArzt achieves the lowest latency (\SI{0.07}{\milli\second}), whereas HMM shows noticeably higher delay (\SI{3.59}{\milli\second}) due to its computational complexity. These results highlight a trade-off between alignment accuracy and runtime efficiency, with LSE and OLTWArzt providing a favorable balance for low-latency use.

The results also show about the characteristics of the datasets.
While the overall alignment performance between (n)ASAP and Batik is comparable, Vienna4x22 shows noticeably higher error variance and kurtosis. This reflects the dataset’s unique structure—22 diverse renditions for each of only four pieces—leading to substantial variability in expressive timing, articulation, and interpretation. These variations present additional challenges for score following and result in heavier-tailed error distributions, as seen in the higher kurtosis values.

% When comparing datasets, Vienna4x22 results in the highest alignment error and lowest AR, consistent with its design that includes 22 highly varied renditions for each piece. This interpretive diversity poses a substantial challenge for score following, especially in a beat-wise evaluation setting where temporal consistency across performers becomes less predictable. In contrast, Batik and (n)ASAP show relatively consistent results, with similar average AAE and AR across both datasets.

\section{Discussions}

\begin{figure}[t]
  \centering
  \centerline{\includegraphics[width=\linewidth]{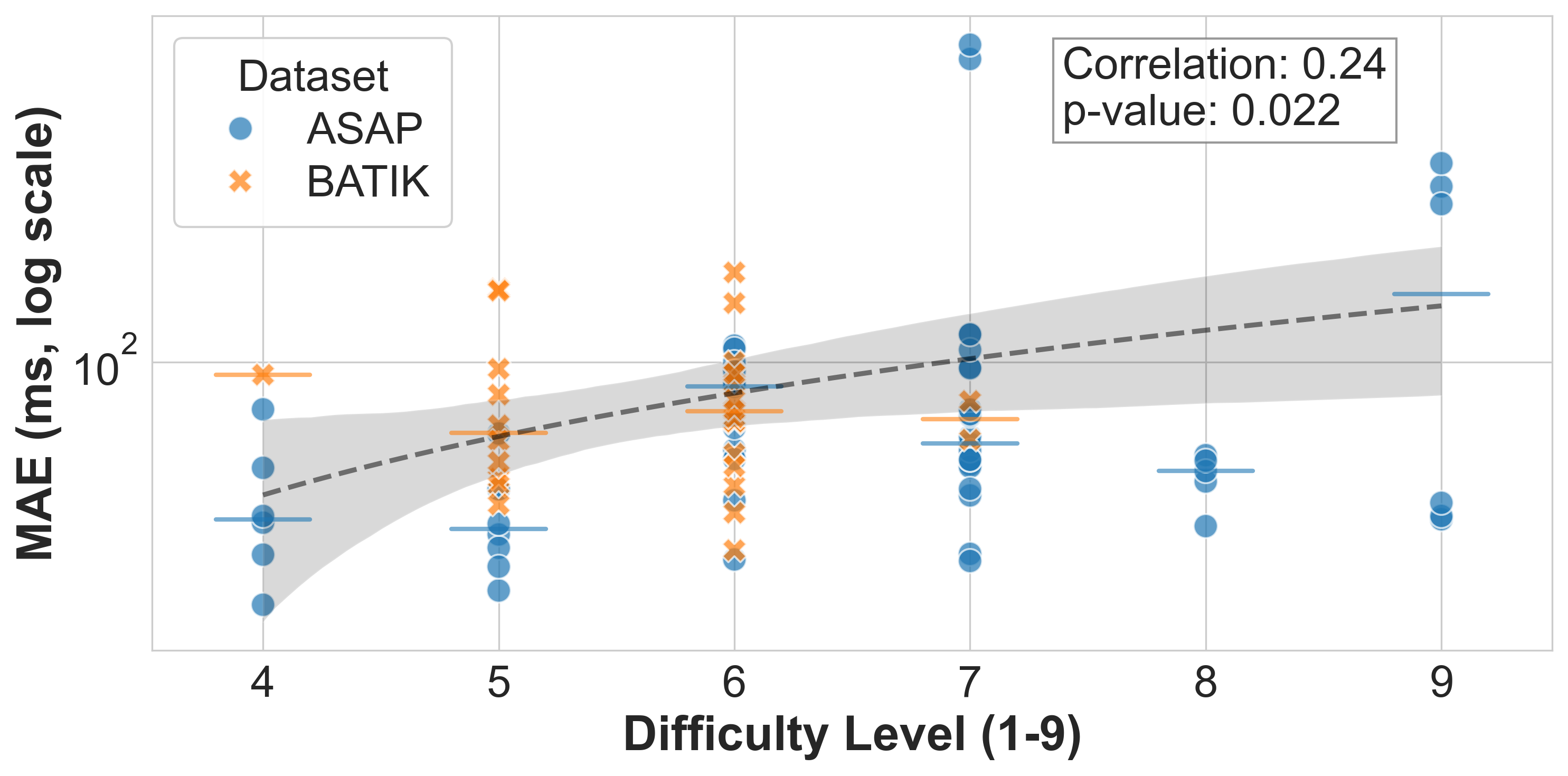}}  
\caption{A scatter plot of mean absolute error (MAE) and Henle's difficulty level in (n)ASAP and Batik dataset. The MAE results are from OLTWArzt.}
\label{fig:scatter-plot}
\end{figure}

Figure~\ref{fig:scatter-plot} further illustrates the relationship between musical difficulty and alignment accuracy for (n)ASAP and Batik. We observe a moderate positive correlation ($r = 0.24$, $p = 0.022$) between MAE and the annotated difficulty levels, indicating that technically more demanding pieces tend to produce larger alignment errors. Vienna4x22 was excluded from this analysis due to its use of short excerpts, which makes consistent difficulty grading unreliable.

\begin{figure}[t]
  \centering
  \centerline{\includegraphics[width=0.98\linewidth]{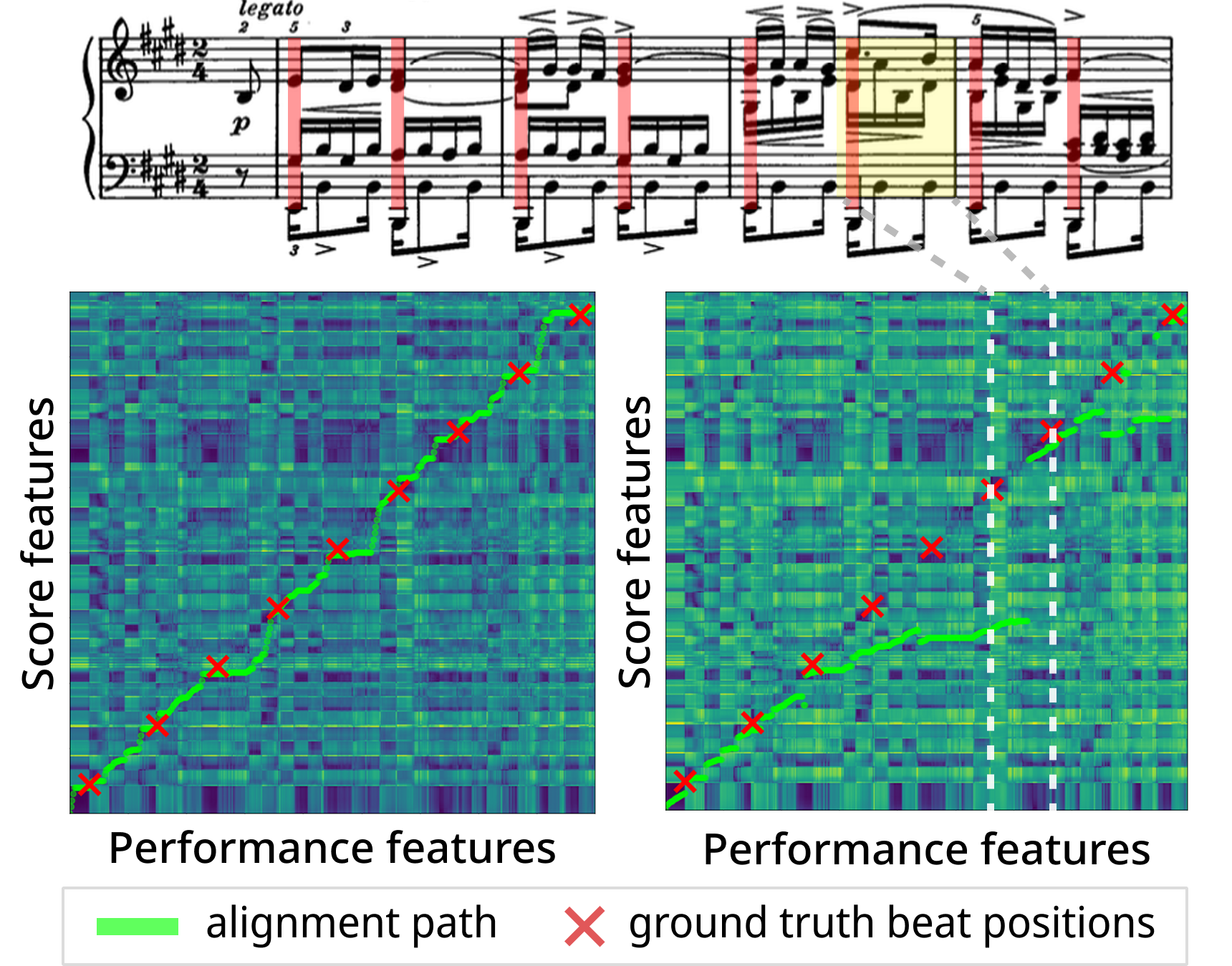}}  
\caption{Two examples of alignment path with beat positions: (left) OLTWArzt, (right) HMM.}
\label{fig:warping-path}
\end{figure}

To further understand how alignment behaviors differ from methods, Figure \ref{fig:warping-path} illustrates an example of alignment result comparing OLTWArzt (left) and HMM (right). 
Although OLTWArzt smoothly follows the beat events, the HMM warping path shows frequent horizontal segments, indicating the ``sticky” tendency to stay near note onsets, reflecting its state-based formulation that emphasizes onset transitions.
This leads to cases where it lingers on sustained notes and becomes locally stuck, showing limited forward momentum.
The corresponding region (highlighted in yellow) exhibits changes in harmony, note density, and dynamics compared to the preceding passage, which provides sufficient contrast for the score follower to recover.

Lastly, we found that not only the choice of evaluation metrics, but also how alignment errors are computed (Section~\ref{sec:mapping-function}) can affect accuracy results to a meaningful extent.
Small differences in error calculation sometimes led to noticeable shifts in reported accuracy.

\section{Use Cases and Applications}\label{sec:applications}

To demonstrate the practicality of our package, we built a lightweight web application that runs locally with real-time audio input or pre-recorded files. 
Built with websocket-based communication, the system responds quickly enough to ensure minimal perceptual delay. Our companion website includes a video demonstration and a link to the source code. 
This application aims to help researchers test their own score following models in an interactive setting. Beyond the web demo, our package is also used as the score following module in the ACCompanion~\cite{cancino-chacon_accompanion_2023}, a real-time accompaniment system. 
These applications demonstrate the versatility of our framework and validate its utility in interactive music scenarios.

\section{Conclusions and Future Work}\label{sec:conclusions}

We presented a systematic framework for real-time audio-based score following as the open-source Python package \emph{Matchmaker}. 
It supports live and simulation-based evaluation with baseline models, enabling reproducible benchmarking across datasets and features. Experiments on three public piano datasets show that the OLTWArzt variant achieves the highest performance and that the onset-sensitive spectral feature (LSE) outperforms chroma in both accuracy and latency.
However, the current framework is limited in its support for tempo models commonly integrated with HMM-based score followers which may partly explain the limited performance of the HMM baseline.
Also, recent works often include learned features or multimodal input which poses a new challenge to evaluate.
Although our evaluation was limited to classical piano, extending \emph{Matchmaker} to other instruments and genres requires only adapting the proper datasets and feature extraction modules.
Future work will extend the framework to support a wider variety of instruments and musical styles, and include additional feature representations, advanced tempo modeling, and multimodal inputs.

% This paper presented a comprehensive overview of real-time audio score following.
% Matchmaker \cite{park_real-time_2024}.

% Recent work on optical score following poses a new challenge to evaluate.

\clearpage

\section{Acknowledgments}
This work has been supported by the Austrian Science Fund (FWF), grant agreement PAT 8820923 (``\emph{Rach3: A Computational Approach to Study Piano Rehearsals}''). Additionally, this work was supported by the National Research Foundation of Korea (NRF) grant funded by the Korea government (MSIT) under Grant RS-2023-NR077289.

%\section{Ethical Statement}

% For BibTeX users:
%\nocite{*}
\bibliography{realtimemusicalignment}

% For non BibTeX users:
%\begin{thebibliography}{citations}
% \bibitem{Author:17}
% E.~Author and B.~Authour, ``The title of the conference paper,'' in {\em Proc.
% of the Int. Society for Music Information Retrieval Conf.}, (Suzhou, China),
% pp.~111--117, 2017.
%
% \bibitem{Someone:10}
% A.~Someone, B.~Someone, and C.~Someone, ``The title of the journal paper,''
%  {\em Journal of New Music Research}, vol.~A, pp.~111--222, September 2010.
%
% \bibitem{Person:20}
% O.~Person, {\em Title of the Book}.
% \newblock Montr\'{e}al, Canada: McGill-Queen's University Press, 2021.
%
% \bibitem{Person:09}
% F.~Person and S.~Person, ``Title of a chapter this book,'' in {\em A Book
% Containing Delightful Chapters} (A.~G. Editor, ed.), pp.~58--102, Tokyo,
% Japan: The Publisher, 2009.
%
%\end{thebibliography}

\end{document}